\documentclass[aps,prb,reprint,superscriptaddress,showpacs]{revtex4-1}
\usepackage{color}
\usepackage{graphicx}
\usepackage[american]{babel}
\usepackage[T1]{fontenc}
\usepackage{amsmath}
\usepackage{amssymb}
\usepackage{amstext}
\usepackage{amsthm}
\usepackage{latexsym}
\usepackage{verbatim}
\usepackage{natbib}
\usepackage{array}
\usepackage{color} 

\usepackage{ulem}   

\def\G{$\Gamma$}

\def\L{\Lambda}

\def\214{Sr$_2$IrO$_4$}

\def\327{Sr$_3$Ir$_2$O$_7$}
\def\J{$J_{1/2}$}
\def\K{$J_{3/2}$}
\def\L{$J_{3/2-3/2}$}
\def\M{$J_{3/2-1/2}$} 

\usepackage[draft]{todonotes}   

\usepackage{xspace}

\begin{document}

\title {Origin of the different electronic structure of Rh- and Ru-doped Sr$_2$IrO$_4$}

\author{V\'{e}ronique Brouet}
\author{Paul Foulquier}
\author{Alex Louat}
\affiliation {Laboratoire de Physique des Solides, CNRS, Univ. Paris-Sud, Universit\'{e} Paris-Saclay, 91405 Orsay Cedex, France}

\author{Fran\c cois Bertran}
\author{Patrick Le F\`evre}
\author{Julien E. Rault}
\affiliation {Synchrotron SOLEIL, L'Orme des Merisiers, Saint-Aubin-BP 48, 91192 Gif sur Yvette, France}

\author{Doroth\'{e}e Colson}

\affiliation{Service de Physique de l'Etat Condens\'{e}, Orme des Merisiers, CEA Saclay, CNRS-URA 2464, 91191 Gif sur Yvette Cedex, France}

\begin{abstract}
One way to induce insulator to metal transitions in the spin-orbit Mott insulator \214 is to substitute iridium with transition metals (Ru, Rh). However, this creates intriguing inhomogeneous metallic states, which cannot be described by a simple doping effect. We detail the electronic structure of the Ru-doped case with angle-resolved photoemission and show that, contrary to Rh, it cannot be connected to the undoped case by a rigid shift. We further identify bands below $E_F$ coexisting with the metallic ones that we assign to non-bonding Ir sites. We rationalize the differences between Rh and Ru by a different hybridization with oxygen, which mediates the coupling to Ir and sensitively affects the effective doping. We argue that the spin-orbit coupling does not control neither the charge transfer nor the transition threshold. 

\end{abstract}

\date{\today}

\maketitle

Inducing metal-insulator transition (MIT) in correlated systems is a major way to reveal new and exotic electronic states \cite{PaschenCondMat20}. After a decade of study of the spin-orbit Mott insulator \214, it has been proved difficult to reach good metallic states, either by doping, beyond the first attempts \cite{GeCaoPRB11}, or by pressure \cite{ChenPRB20}. Substitutions of Ir with $4d$ transition metals (TM) induces a metallic state, but also raises many question about the role of disorder. Which is the main driving force of the MIT, either reduced spin-orbit coupling (SOC)\cite{LeeTokuraPRB12} or effective doping \cite{ClancyPRB14} has recently been challenged again by an ARPES study in favor of SOC \cite{ZwartsenbergNatPhys20}. XAS \cite{ClancyPRB14,ChikaraPRB17} and ARPES \cite{CaoDessauNatCom16,LouatPRB18} have shown that, unexpectedly, Rh, isovalent to Ir, dopes holes into \214, as if its energy levels were below those of Ir (see Fig. 1). On the contrary, Ru, which has one more hole than Ir, does not seem to dope at low values \cite{CavaPRB94,DhitalNatCom14}, suggesting an opposite hierarchy between energy levels. In this situation, Ru should transfer electrons to Ir, but this is forbidden by the Coulomb repulsion on Ir, as long as the insulating Mott state resists. To complicate things further, the atomic Coulomb repulsion U is at least as strong for Ru and Rh as for Ir and the metallic states found in Sr$_2$RuO$_4$ and Sr$_2$RhO$_4$ are only understood by their smaller SOC that preserves the degeneracy of the conduction band, which reduces the impact of correlations\cite{BJKimPRL08,MartinsPRL11}. Hence, there is a strong interdependence between the possibility of charge transfer on Ir, the existence of a Mott state and the effective value of SOC, which makes the problem highly non-trivial, as all these parameters may change through doping. 

Sr$_2$Ir$_{1-x}$Ru$_x$O$_4$ remains isostructural up to x=0.55, with a slight decrease of the in-plane rotation of the oxygen octahedra \cite{YuanPRB15}. An insulator to metal transition was observed around x$\simeq$0.4 by transport \cite{YuanPRB15}, in concomittance with the disappearance of the long range magnetic order \cite{CalderPRB15}. A similar behavior was observed in Ru-doped Sr$_3$Ir$_2$O$_7$ \cite{DhitalNatCom14,WangPNAS18}. The substitution value to reach MIT is much larger for Ru than Rh (x$\simeq$0.1) and was suggested to correspond to percolation of Ru-rich metallic puddles \cite{DhitalNatCom14,FengPRE08}. Nevertheless, the only ARPES study available to date reveals a Fermi Surface (FS) containing 5-x electrons at x=0.4, as if a simple hole doping has been reached\cite{ZwartsenbergNatPhys20}. How charge transfer emerges from the phase separation at early dopings has not been explained yet. 

\begin{figure}
\centering
\includegraphics[width=0.9\columnwidth]{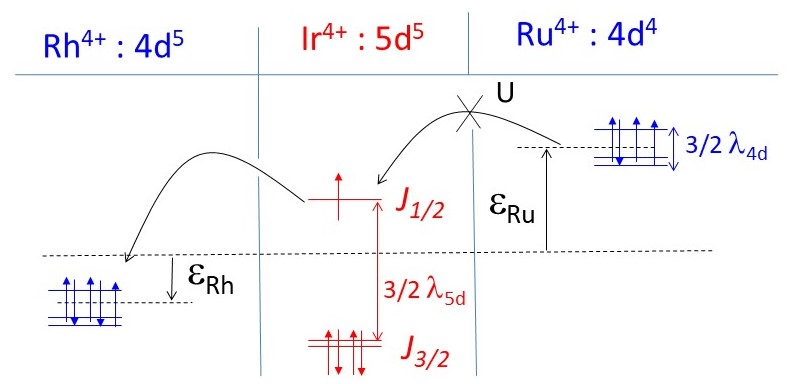}
\caption{Sketch of the ionic t$_{2g}$~levels for Rh, Ir and Ru. SOC splits them into one \J~and two \K~levels ({$J_{3/2-m_J}$ with $m_J=\pm1/2,3/2$}), with a much larger value for Ir, being a $5d$ TM, than Rh and Ru. We assume a shift $\varepsilon$ between Ir and the other TM (see Fig. \ref{Hybridation} for a discussion of its origin). Electrons should be transferred to the lowest available energy level, unless the Coulomb repulsion U forbids double occupation.} 
\label{Sketch}
\end{figure}
\begin{figure}
\centering
\includegraphics[width=1\columnwidth]{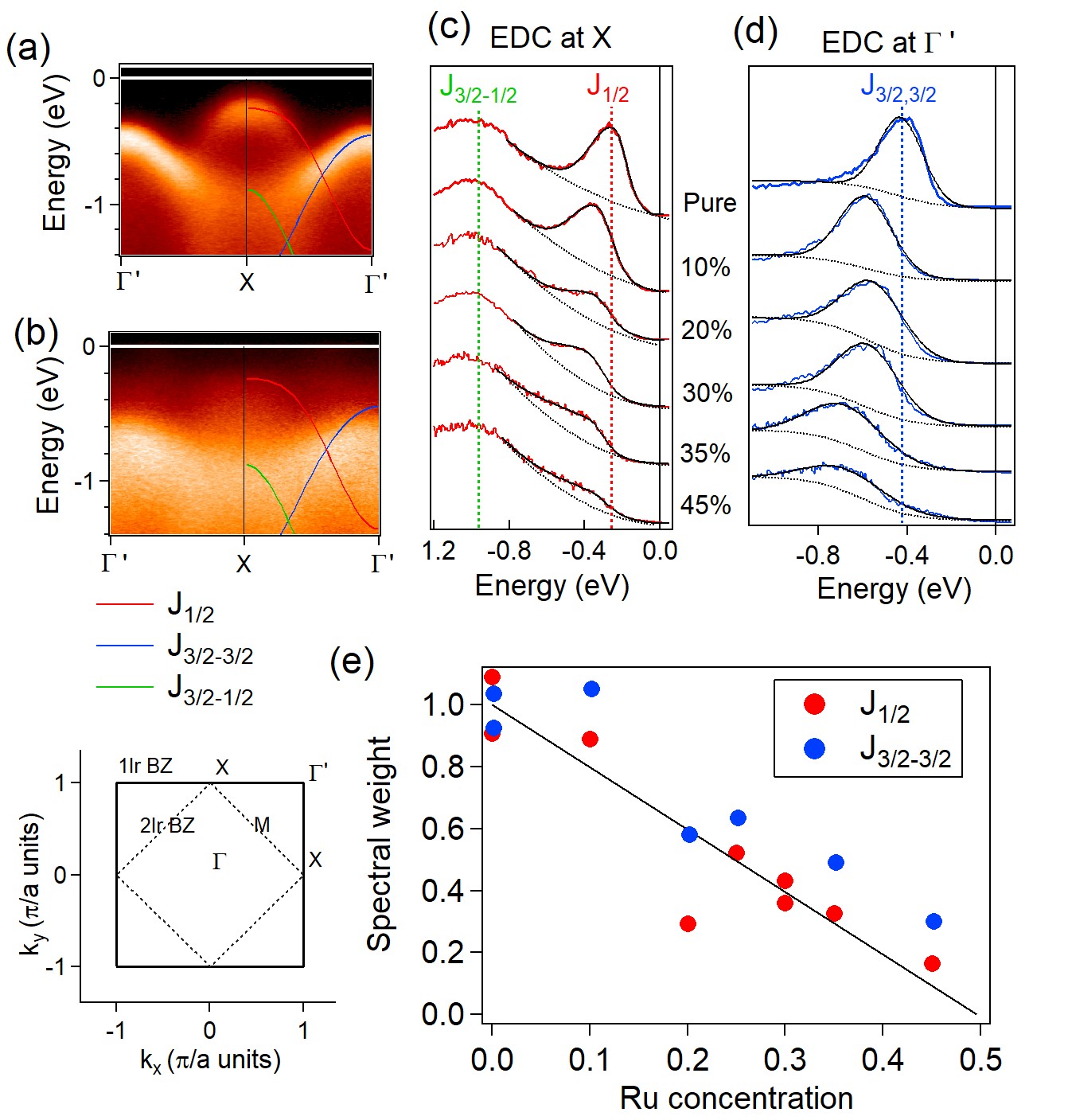}
\caption{(a) Energy-momentum plot along $\Gamma$'X in \214 ($k_y$=3, see BZ sketch below). The lines highlight the three main bands \cite{sup}. (b) Same for \214 doped with 45\%~Ru. (c) EDC at X as a function of Ru doping (as indicated) fitted with a polynomial background (dashed line) and an asymmetric gaussian (black line). (b) Same at $\Gamma$' fitted with a fixed step-like background and a gaussian. (e) Relative spectral weight of the \J~and \L~peaks compared to the pure, when spectra are normalised to the background intensity. More samples are included than those shown in (c-d).} 
\label{EDC}
\end{figure}

We report the evolution of the electronic structure as a function of Ru substitutions with ARPES and detail the Fermi Surface (FS) at x=0.45. We show that the MIT is not due to a shift of the \214 bands towards the Fermi level, as was observed for Rh doping\cite{CaoDessauNatCom16,LouatPRB18}. Instead, the \214 bands gradually lose weight and a new set of bands appear near the Fermi level, which effective SOC is much smaller than in the Rh case. This rules out that SOC is the key factor of the MIT. Supported by DFT calculations, we explain the different SOC by the dominant Ru character of the bands near $E_F$, which contrasts with the dominant Ir character of the bands near $E_F$ for Rh doping. We assign this difference to a different hybridization with  oxygen.

The samples were prepared using a self-flux method, as reported in ref. \onlinecite{KimScience09}. Their exact doping was estimated by Energy Dispersion X-ray analysis and the structure checked by single-crystal X-ray diffraction. ARPES experiments were carried out at the CASSIOPEE beamline of SOLEIL synchrotron, with a SCIENTA R-4000 analyser, 100 eV photon energy and an overall resolution better than 15meV.

\vspace{0.2cm}

Fig. \ref{EDC}a gives an ARPES view of the bands in the pure compound. The \J~band peaks at X (red line) and the two \K~bands, respectively at \G~for \L~(blue line) and X for \M~(green line) (these notations are defined in Fig. 1). For clarity, we only indicate the bands with large ARPES weight in these experimental conditions\cite{sup}. In Fig. \ref{EDC}(c-d), we show the Energy Distribution Curve (EDC) at X and $\Gamma$', respectively. The peaks only slightly move to higher binding energies, which is completely different from the case of Rh, where all peaks rigidly move to the Fermi level \cite{LouatPRB18}.  

The peaks also seem to broaden and/or lose weight. This is clearer for \J, which intensity can be directly compared to the one of the filled \M. We use it as a background reference (dashed line) and extract the \J~peak spectral weight by fitting the remaining peak with an asymmetric gaussian. The area normalised to this background is reported in Fig. \ref{EDC}d and is consistent with a linear decrease as 1-2$x$. The intensity of \L~is more difficult to evaluate, because its background is not as well defined. The image in Fig. \ref{EDC}b shows that, at 45\%~Ru, its intensity has indeed weakened, as it became comparable to that of the \M. Assuming a step-like background, we obtain a similar decrease of intensity as \J~in Fig. \ref{EDC}e. 

In a correlated system, it would be natural to find a loss of intensity corresponding to a transfer of spectral weight from an incoherent Hubbard-like band to a coherent band near the Fermi level. However, doping a half-filled band, one would rather expect a (1-$x$) dependence for the incoherent part weight at small doping $x$ \cite{GeorgesRMP96}. More puzzingly, no change would be expected for the \L~band, which is completely filled. 

\begin{figure}
\centering
\includegraphics[width=1\linewidth]{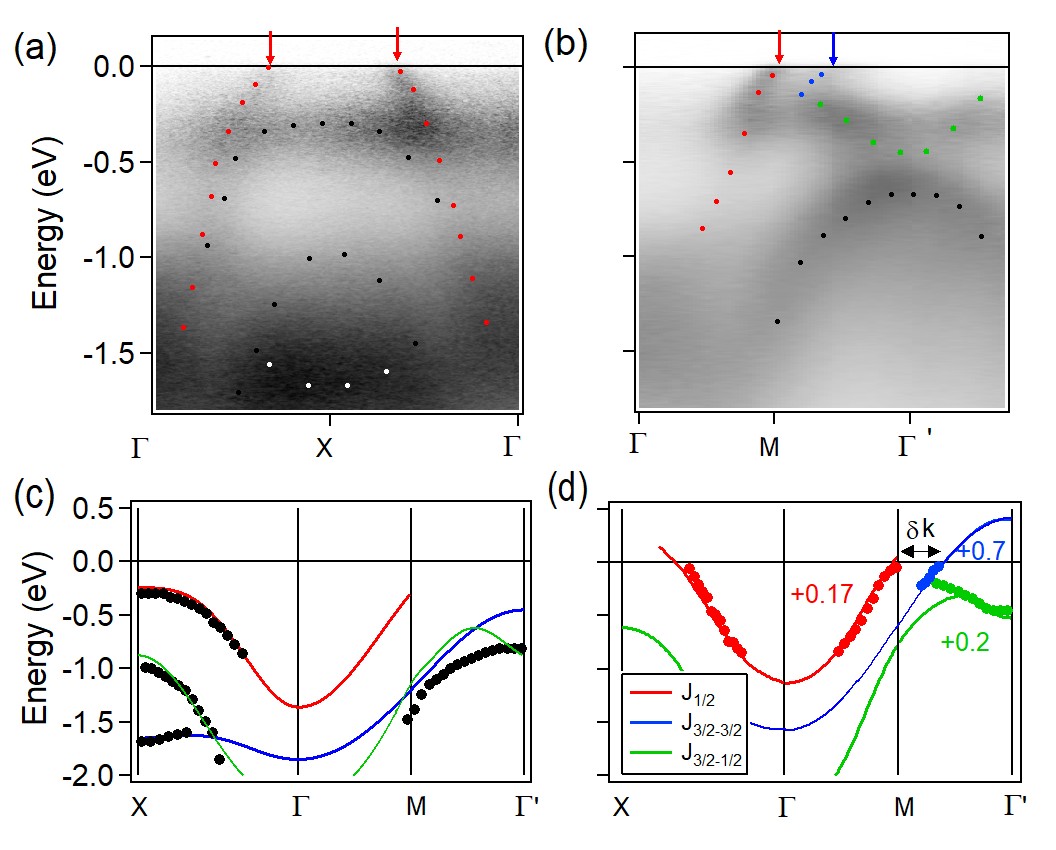}
\caption{Energy-momentum plot of ARPES intensity measured at 20K with 100eV photon energy and Linear Horizontal polarisation, along (a) $\Gamma$X ($k_y$=2) and (b) $\Gamma$M . To enhance the low features near $E_F$, the image is multiplied by a Fermi step along y, centered at -0.27eV with width 0.2eV and amplitude 20. Markers are guide for the eyes of the different bands. (c) Comparaison of the high binding energy bands (black markers) with the dispersion measured in pure \214 (lines). The dispersion are extracted either from MDC fits or from local maximum. (d) Comparaison of the bands near $E_F$ (color markers) with the dispersion measured in 15\% doped Rh \cite{LouatPRB18}. The Rh dipsersions are shifted up to match the Ru data by the indicated amounts. } 
\label{ModelDisp}
\end{figure}

In Fig. \ref{ModelDisp}(a-b), we further note as black and white markers the bands corresponding to these peaks in k-space, along $\Gamma$X and $\Gamma$M. These markers are reported in Fig. \ref{ModelDisp}(c) and compared to the dispersion measured by ARPES in the pure compound \cite{LouatPRB19}, to which they are nearly identical. On the other hand, there are three new bands appearing closer to the Fermi level, emphasized as color markers. Two of them cross the Fermi level, as indicated by arrows. They exhibit weak QP peaks but no pseudogap \cite{sup}. The resulting FS \cite{sup} is similar to that observed in ref. \onlinecite{ZwartsenbergNatPhys20}.

The dispersions of these metallic bands, reported in Fig. \ref{ModelDisp}(d), correspond quite well to the expectation for three $t_{2g}$ bands split by SOC. Especially, the shape of the green band forming an electron-like pocket centered at \G' is typical of the avoided crossing between $d_{xy}$ and $d_{x2-y2}$, observed for \M~in compounds where the oxygen octahedra are rotated \cite{KimPRL06}. This band does not seem to reach the Fermi level. The red band forms a large squarish electron pocket around $\Gamma$, containing $n$=0.8 electrons according to the Luttinger theorem, while the blue band forms smaller squarish pockets around $\Gamma$', containing $n$=0.18 holes \cite{sup}. This FS structure looks like the $\alpha$ and $\beta$ sheets observed in Sr$_2$RuO$_4$ \cite{DamascelliPRL00}. Adding electrons of these three bands, we obtain a total $n=$4.62 electrons, remarkably close to the 5-x electrons expected for a simple hole doping by Ru. This implies that, despite the coexistence of two sets of bands (insulating-like in black and metallic-like in color), the FS does not correspond to a phase separation between electronically isolated Ir and Ru clusters.

To further characterize the metallic bands, we compare them in Fig.~\ref{ModelDisp}(d) with dispersions measured for 15\% Rh \cite{LouatPRB19}. These models describe well the Ru-doped dispersions, implying there is no significant renormalization. Similarly, there is no significant sharpening of the peak near $E_F$. From this point of view, Ru-doped \214 is similar to other doped \214, lacking the traditional fingerprints of a correlated Fermi liquid, contrary to Sr$_2$RhO$_4$ \cite{BaumbergerPRL06,KwonPRL19} and Sr$_2$RuO$_4$ \cite{TamaiPRX19}, where renormalizations of factors 2-3 are observed near $E_F$.  

For \J, the Rh model has to be shifted up by 170~meV, which can be understood from the different filling (4.55 and 4.85, respectively). However, a huge shift of 0.7eV is needed for the \L~band, which is still 0.2eV below $E_F$ at \G~for $x_{Rh}$=0.15. This indicates a highly non-rigid shift between Ru and Rh doping and suggests a drastic reduction of the SOC, which controls the splitting between \J~and \K. Indeed, the momentum splitting $\delta$k=0.27$\pi$/a between the two bands at $E_F$ is similar to the one measured in the purely $4d$ Sr$_2$RhO$_4$ \cite{BaumbergerPRL06,KimPRL06}. Extracting a SOC value from the dispersions is however difficult, as it could be renormalized \cite{TamaiPRX19} or enhanced \cite{LiuPRL08,ZhouPRX18} by correlations and also affected by a different closure of the Mott gap in the two cases. Assuming that SOC is a simple average between Ir and the TM dopant, one would expect a 30\% stronger reduction for doped Ru for which $x$ is larger, but this hardly explains a value appearing similar to $4d$ metals. A better evaluation would take into account the atomic weight of the bands. Following the hierarchy of energy levels sketched in Fig. 1, one can expect the top of the band to have more Ru character, hence a SOC more effectively reduced than expected from $x$ and the opposite for Rh doping \cite{sup}. The different alignment of energy levels (i.e. on-site energies $\varepsilon$) then gives a qualitative explanation for the difference in SOC in the FS near the MIT. 

\vspace{0.2cm}

\begin{figure}
\centering
\includegraphics[width=1\linewidth]{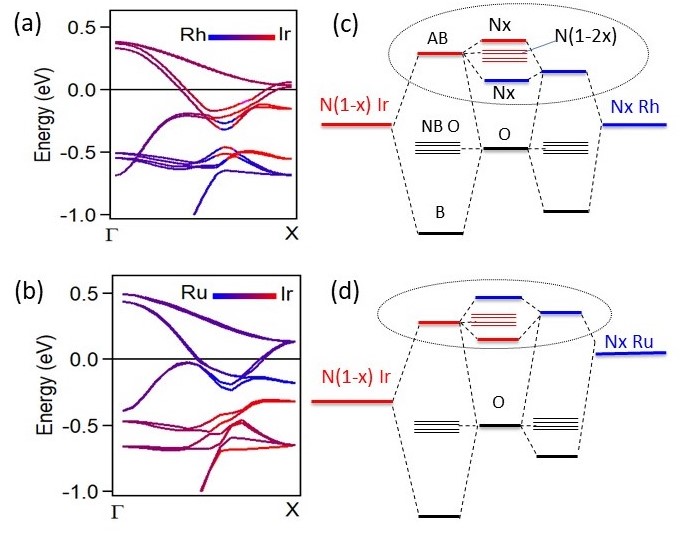}
\caption{(a) Band structure calculated for an ordered structure Sr$_2$(IrRh)$_{0.5}$O$_4$ along $\Gamma$X, without SOC. The color is proportional to the atomic character (red-blue scale). (b) Same for Ru. (c-d) Sketch of the energy levels of oxygen, Ir and TM dopant and their relative hybridization, based on calculation presented in supplementary, but not to scale. The hybridized states have mixed atomic character, their color indicates the dominant one. The circled part represents hybridization of N Ir and TM levels. } 
\label{Hybridation}
\end{figure}

We now consider possible origins for these different $\varepsilon$. It has been proposed that the smaller SOC leads to electron trapping in Rh \cite{CaoDessauNatCom16,LiuFranchiniPRB16}, but Ru does not trap electrons, so it cannot be the only reason. Alternatively, a positive impurity potential was assumed for Ru in ref. \onlinecite{ZwartsenbergNatPhys20}, because of its different charge, but this does not explain why Rh hole dopes. To get a qualitative idea of how the energy levels could align, we performed DFT calculations for the simplest structure mixing the two atoms, an ordered \214 structure with 50\% Ir replaced by another TM. As shown in Fig. \ref{Hybridation}, the distribution of Ir and dopant weight is strikingly different, with more Ru weight on top of the band and more Rh weight at the bottom. The respective contribution are of the order 40\%-60\% at the Fermi level. As SOC is not included here, this suggests that the origin of the difference is rooted in basic properties of the electronic structure. 

To some extent, mixing different TM in a compound reproduces locally what happens at oxides heterostructures, where charge transfer is commonly observed as a result of different electronegativity \cite{ChenMillisJPCM17} or hybridization strength \cite{GrisoliaNatPhys16}. Observing different valence states of TM doped in oxides is actually not so uncommon \cite{StreltsovJPCM16,HossainPRL08}. 
As described in ref.~\onlinecite{ChenMillisJPCM17}, to "align" the energy levels, a natural reference is the oxygen states, which must be shared between the two TM. Following this idea, we present in supplementary DFT calculations to evaluate the coupling with oxygen\cite{sup}. The essential results are sketched in Fig. \ref{Hybridation}(c-d). Hybridization between oxygen and TM creates antibonding (AB) and bonding (B) states, respectively dominated by the TM and the oxygen, as well as non-bonding (NB) states for oxygens states without TM partners. Their splitting depends both on the coupling strength and the relative initial energy of TM and oxygen \cite{sup} and turns out to be significantly smaller for Rh than Ir (4.7eV vs 5.4eV), as could be anticipated from the smaller extension of $4d$ orbitals. This would create an energy difference between Ir and Rh AB levels initially absent. Having one less electron, Ru displays a smaller electronegativity, which destabilizes its initial energy level compared to O. The calculation suggests this effect nearly compensates the smaller coupling strength of $4d$ element and could reverse the respective positions of AB levels.

In the circled part of Fig. \ref{Hybridation}, we consider the hybridization of $N$ atoms including $x$ TM, starting from these relative positions. This shall create $Nx$ B and $Nx$ AB states with larger atomic character from the closest energy level. The difference in relative position for Ru and Rh induced by coupling to oxygen explains qualitatively the different distribution of atomic character in Fig. \ref{Hybridation}(a-b) and, by extrapolation, the tendency of isolated Rh to trap one electron. If there are less Rh or Ru atoms than Ir ($x\ll$0.5), there shall be a corresponding number $N(1-2x$) of unpaired Ir NB states. As this is exactly the weight we found for the peaks remaining at Ir positions in Fig. \ref{EDC}, it is tempting to identify them with NB-like Ir states. It is remarkable that they remain insulating-like, despite the progressive formation of a metallic environment. A full description of the new band structure, taking correlation effects into account, is beyond the scope of this paper, but should be very interesting. We note, for example, that in Fig. \ref{ModelDisp}, the NB \J~could form a nearly flat band rather than follow the original \214 dispersion. The existence of NB states could also solve the puzzle of the pseudogap observed in Rh-doped metallic state \cite{CaoDessauNatCom16,LouatPRB18}. We have shown that the pseudogap is not restricted to the region near $k_F$, but is on the contrary clearest at X, where only incoherent weight is expected \cite{LouatPRB19}. This becomes natural if the pseudogap is due to a distorsion of the lineshape near $E_F$, created by the underlying structure of NB states, that would be for Rh close to the metallic band, but possibly remaining distinct. 

Recently, Zwartsenberg et al. argued that the MIT occurs when the effective SOC reaches a certain threshold ($\lambda$=0.44eV) and that it is obtained at a larger $x$ for Ru than Rh, defining a different substitution threshold for the MIT in each case \cite{ZwartsenbergNatPhys20}. This seems to be in strong contradiction with our finding that the two metallic states emerge with very different effective SOC, and moreover, much smaller for Ru than Rh. The problem is that SOC is modulated across the band structure depending on its atomic content \cite{sup}. Upon Ru doping, it is smaller near $E_F$, as we found for the metallic bands in this paper, and larger at higher binding energy where Ir dominates, as ref. \onlinecite{ZwartsenbergNatPhys20} estimated from ARPES relative orbital intensities. These different estimations are not in contradiction and they are in fact based on the same idea of the influence of the on-site energies $\varepsilon$ on the dilution of the effective SOC. Now, regarding the MIT, the common wisdom is that the role of SOC is simply to lift the degeneracy \cite{BJKimPRL08,MartinsPRL11}. In this respect, whatever the precise SOC value is, the Rh and Ru MIT do not happen at the same effective degeneracy. Therefore, we do not see how the MIT could be controlled by SOC. 

In our scenario, the reason for the different doping threshold depends on the way holes are introduced, either in the lower Hubbard band (Rh case) or in AB states generated by the hybridization between Ir and TM (Ru case). The resulting metallic states are very different and it is therefore not surprising that the MIT does not take place at the same doping. In the Rh case, we have even shown recently that holes and electrons coexist near the MIT \cite{Fruchter21}. In the Ru case, it is necessary to create enough AB states to develop metallicity, which implies having a doping near x=0.5. 

To conclude, the electronic structures upon Rh and Ru dopings differ by much more than a different degree of hole doping, they cannot be deducd from each other by a rigid shift. Despite this, the created metals both have a low degree of coherence, as evaluated from the absence of renormalization and no well behaved quasiparticle peaks. This appears a characteristic of metallic iridates. This study further gives a vivid example of how carriers can be trapped or created at different sites in iridates. We argue that this is due to differences in energy levels arising from different local hybridization with oxygen, which may play a particularly important role for $5d$ systems. This may reorient our way to think about these materials, as similar effects could be expected around oxygen defects (vacancy, local distorsion) or dopants and be crucial to understand doping.  

\vspace{0.2cm}

We acknowledge interesting discussions with Cyril Martins, S\'ebastien Burdin and Andrea Damascelli. This work was supported by the Agence Nationale de la Recherche grant "SOCRATE" (ANR-15-CE30-0009-01). 

\bibliography{Biblio_Ru}


\begin{widetext}
\newpage
\Large
\textbf{Supplementary Information}

\vspace{0.4cm}

\noindent\normalsize

\noindent
\textbf{\214 band structure}

For convenience, we recall in Fig. \ref{Sup_Bands}a a typical band structure for \214 calculated using the Wien2k software \cite{Wien2k}. Similar results can be found in many papers \cite{MartinsPRL11}. We overlay with color lines the bands we call \J~(red), \L~(blue) and \M~(green) in the maintext, from their dominant orbital character. We also show as dotted lines the bands folded into the 2Ir BZ, typically having small ARPES weight (see ref. \onlinecite{LouatPRB19} for details). For example, $\Gamma$ and $\Gamma$' are equivalent in the 2Ir BZ, but the electronic structure appears very different at these two points in Fig. 3b. Similarly, the structure along $\Gamma$X appears very different at $k_y$=2 (Fig. 3) and $k_y$=3 (Fig. 2), because \J~is a "main" band for the former and a "folded" band for the later. Considering only the main bands, we obtain in Fig. \ref{Sup_Bands}b a simpler band structure, which is the one we use as reference in the text. All markers shown in Fig. 3 correspond to such "main" bands. 

At $\Gamma$', the \J~and \L~bands nearly correspond to the $d_{xz}$ and $d_{yz}$ doublet split by SOC. Therefore, the distance noted $\delta$E would be a good measure of the SOC strength, but it is not accessible in ARPES, as it occurs above $E_F$. The distance $\delta$k also directly depends on the SOC strength, but it also depends on the band dispersion. 

As it is well known, and reproduced by LDA+U calculations\cite{ZhouPRX18}, the effective SOC in \214 is larger than in this calculation, which brings the \L~band entirely below the Fermi level. This also means that the bands do not cross where it is predicted in the calculation, which is the reason why we ommit the hybridization gaps in our models (they are indeed absent in ARPES). 

\begin{figure}[h]
\centering
\includegraphics[width=0.75\linewidth]{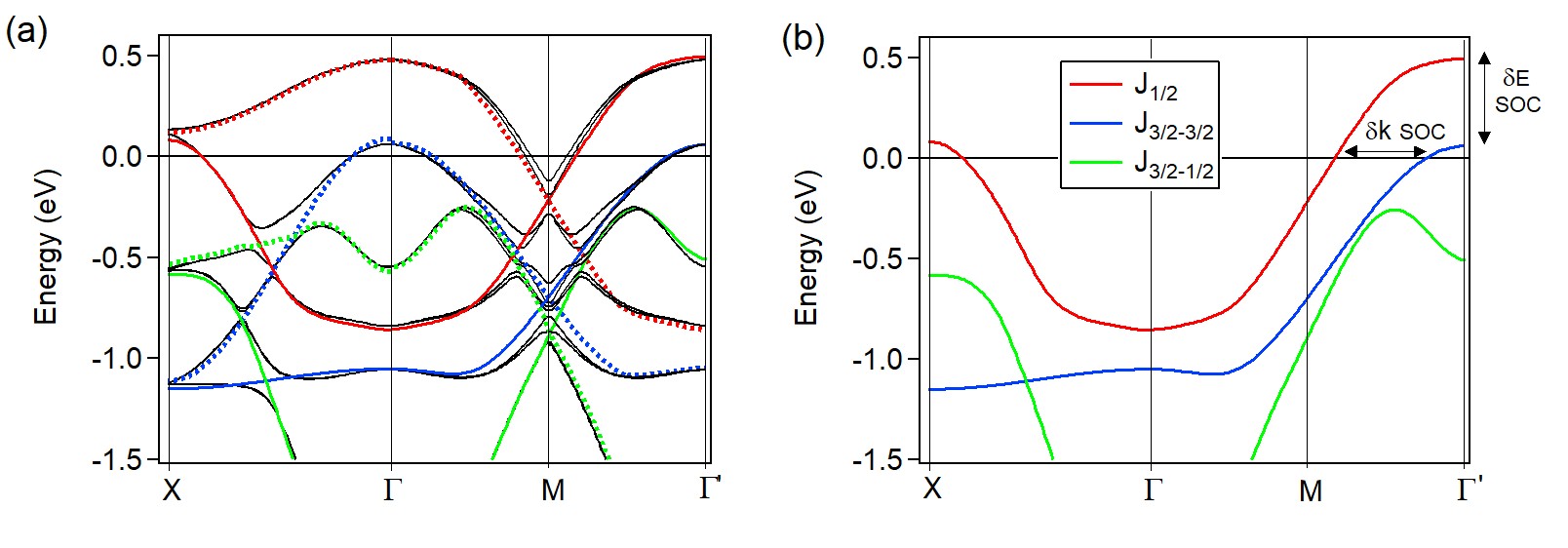}
\caption{(a) Band structure calculated for \214 (black lines). Color lines are superimposed to describe the main bands, in red for \J, blue for \L~and green for \M. The bands that can be considered as "folded" are dotted lines. (b) Plot of the fit of the main bands, which gives a simplified view of the structure, convenient to compare to ARPES measurements. } 
\label{Sup_Bands}
\end{figure}

\vspace{0.5cm}
\noindent
\textbf{Fermi Surface ofr \214 doped with 45\%Ru}

\vspace{0.2cm}

The Fermi Surface of \214 doped with 45\% Ru is presented in Fig. \ref{Sup_FS}a. It consists of a large squarish pocket around $\Gamma$ (red line) and smaller pockets around $\Gamma$' (blue lines). It is very similar to the one measured at x=0.4 in ref. \onlinecite{ZwartsenbergNatPhys20}, although it was rather described by hole pockets around X instead of electron pocket around $\Gamma$in this reference. We note that the folded sheets expected with respect to the dotted black lines (2Ir BZ) have a very low intensity, indicating a weakening of the role of the structural distorsion (i.e. the rotation of the oxygen octahedra). The distorsion is however still present as proved by the shape of the \M~band in Fig. 3 (see main text). 

In Fig.~\ref{Sup_FS}b, selected lineshapes at $k_F$ are presented. We observe a small QP peak along $\Gamma$M, weak but similar to the La-doped case \cite{BrouetPRB15,DeLaTorrePRL15}. It further weakens towards $\Gamma$X, independently of experimental conditions, but without shifting. Within a 10-20meV incertitude due to the low peak intensity, there is no pseudogap, contrary to the Rh case \cite{CaoDessauNatCom16,LouatPRB18}. The higher intensity along $\Gamma$M is rather unexpected as the hole pockets start to develop from $\Gamma$X and we do not have an explanation for this at the moment.

The Luttinger theorem relates the area of the FS sheet (assuming a 2D structure) to the number of carriers it contains. For a square pocket $n=2k_F^2$, with $k_F$ in $\pi$/a units, yielding for the red pocket with $k_F$=0.63(3)$\pi$/a, n=0.80(7) electrons. For the blue hole-pocket, $k_F$=0.3(5)$\pi$/a, giving $n$=0.18(5) holes. 

Fig.~\ref{Sup_FS}c compares this FS with those observed in related compounds. This structure is very close to the $\alpha$ and $\beta$ sheets observed in Sr$_2$RuO$_4$ (red lines) \cite{DamascelliPRL00}. The $\gamma$ sheet is missing due to the oxygen rotation. Without SOC, $\alpha$ and $\beta$ are built by $d_{xz}$ and $d_{yz}$, $\gamma$ by $d_{xy}$. Including SOC, they become \J, \L~and \M. The overall structure observed in Sr$_2$RhO$_4$ (green lines) is similar to our case, but with a higher electron filling (n=5 instead of 4.5). The intensity of the folded FS is also higher \cite{KimPRL06}. In \214 doped with 15\%Rh \cite{LouatPRB18} (light blue), the FS consists in hole pockets around X, because of a residual gap at M. The \L~band ($\beta$ sheet) is still enirely below $E_F$.

\begin{figure}[h]
\centering
\includegraphics[width=0.75\linewidth]{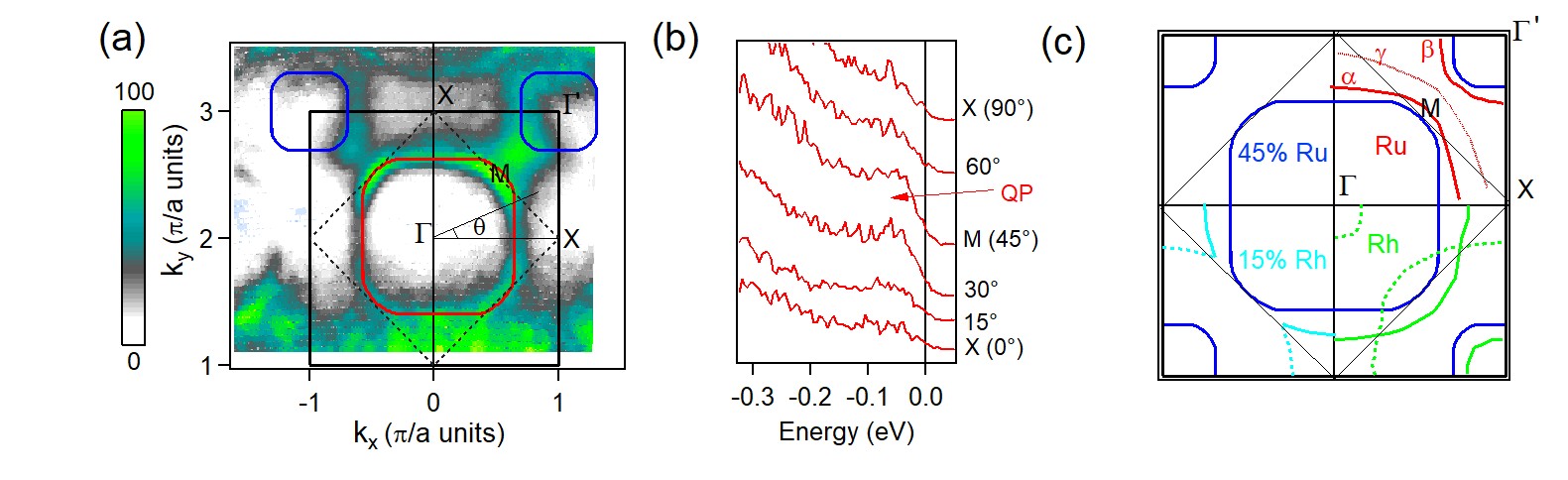}
\caption{(a) Fermi Surface measured at 20K with 100eV photon energy and linear horizontal polarization for \214 doped with 45\% Ru. The black square is the 1Ir BZ (of side a) and the dashed one the 2Ir BZ. (b) EDC measured at k$_F$ along the red pocket for the indicated angle $\theta$. (c) Comparaison of FS in different compounds : \214 doped with 45\% Ru (blue), \214 doped with 15\% Rh (light blue), Sr$_2$RuO$_4$ (red) and Sr$_2$RhO$_4$ (green). The three FS sheets traditionally noted $\alpha$, $\beta$ and $\gamma$ are indicated.} 
\label{Sup_FS}
\end{figure}

\vspace{0.5cm}
\noindent
\textbf{Tight-binding model of the effective SOC for two hybridized atoms}

\vspace{0.2cm}

We first consider two atoms with different energy levels -$\epsilon$ and $\epsilon$. We calculate their hybridization with a constant interaction $V$, as shown in Fig. \ref{SupTB}a by black levels, solving a simple two-level system. The hybridized levels : 

\noindent |$\psi^{\pm}$>=a|1> + b|2> have energies $E^{\pm}=\pm \sqrt{\varepsilon ^{2}+V^{2}}$, with a/b=V/($E^{\pm}$+$\varepsilon$). 

\noindent Hence, the splitting depends both on V and $\varepsilon$ and the hybridized levels have a dominant character of the closest atomic levels. For $\varepsilon$=0, these are the usual bonding (B) and antibonding (AB) states |$\psi^{\pm}$>=|1> ${\pm}$~|2>. 

\vspace{0.15cm}

To include SOC, we consider two orbitals $d_{xz}$ and $d_{yz}$. The interactions are described by the following hamiltonian in a basis {$d_{xz1}$, $d_{yz1}$, $d_{xz2}$, $d_{yz2}$}. Spin must be considered when SOC is included, which adds another block that is the hermitian conjugate. 
\begin{equation}
\mathcal{H}=\left[\begin{matrix}\begin{matrix}-\varepsilon&i\lambda_1/2\\-i\lambda_1/2&-\varepsilon\\\end{matrix}&\begin{matrix}V\ \ \ \ \ \ \ &0\\0\ \ \ \ \ \ \ &V\\\end{matrix}\\\begin{matrix}V\ \ \ \ \ \ \ &0\\0\ \ \ \ \ \ \ &V\\\end{matrix}&\begin{matrix}\varepsilon&i\lambda_2/2\\-i\lambda_2/2&\varepsilon\\\end{matrix}\\\end{matrix}\right]
\end{equation}

We suppose the two atoms have very different SOC parameters $\lambda_1$ and $\lambda_2$. We define an effective SOC by the splitting of the hybridized levels. We find that this effective SOC is just the average value of the two $\lambda$ weighted by the atomic weights. In our example (Fig.\ref{SupTB}b), it is smaller than $\lambda_1$ on the top levels, dominated by the second atom with small $\lambda$, and larger at the bottom, where the situation is reversed. 

\vspace{0.15cm}

In Fig. \ref{SupTB}c, we include a cosine dispersion along $k_x$ for $d_{xz}$ [V(k) = V cos($k_x$)] and $k_y$ for $d_{yz}$. This describes well iridates, as the $xy$ band, which is completely filled, plays only a marginal role. Without SOC and without $\varepsilon$, the bonding bands (red and blue lines) are the usual $d_{xz}$ and $d_{yz}$ ("main bands" of the 2IrBZ). The antibonding bands (dotted lines) are the folded bands of the 2Ir unit cell \cite{BrouetPRB12}. When we consider $\varepsilon$, a gap 2$\varepsilon$ opens where the bands cross. This is an artefact of the ordered structure and does not play a role in our model. 

When SOC is included (Fig. \ref{SupTB}d), a gap opens where $d_{xz}$ and $d_{yz}$ are degenerate, which happens at $\Gamma$. As in the case without dispersion, the top of the band is dominated by the small $\lambda$ and the bottom by the larger $\lambda$.

\begin{figure}
\centering
\includegraphics[width=0.75\linewidth]{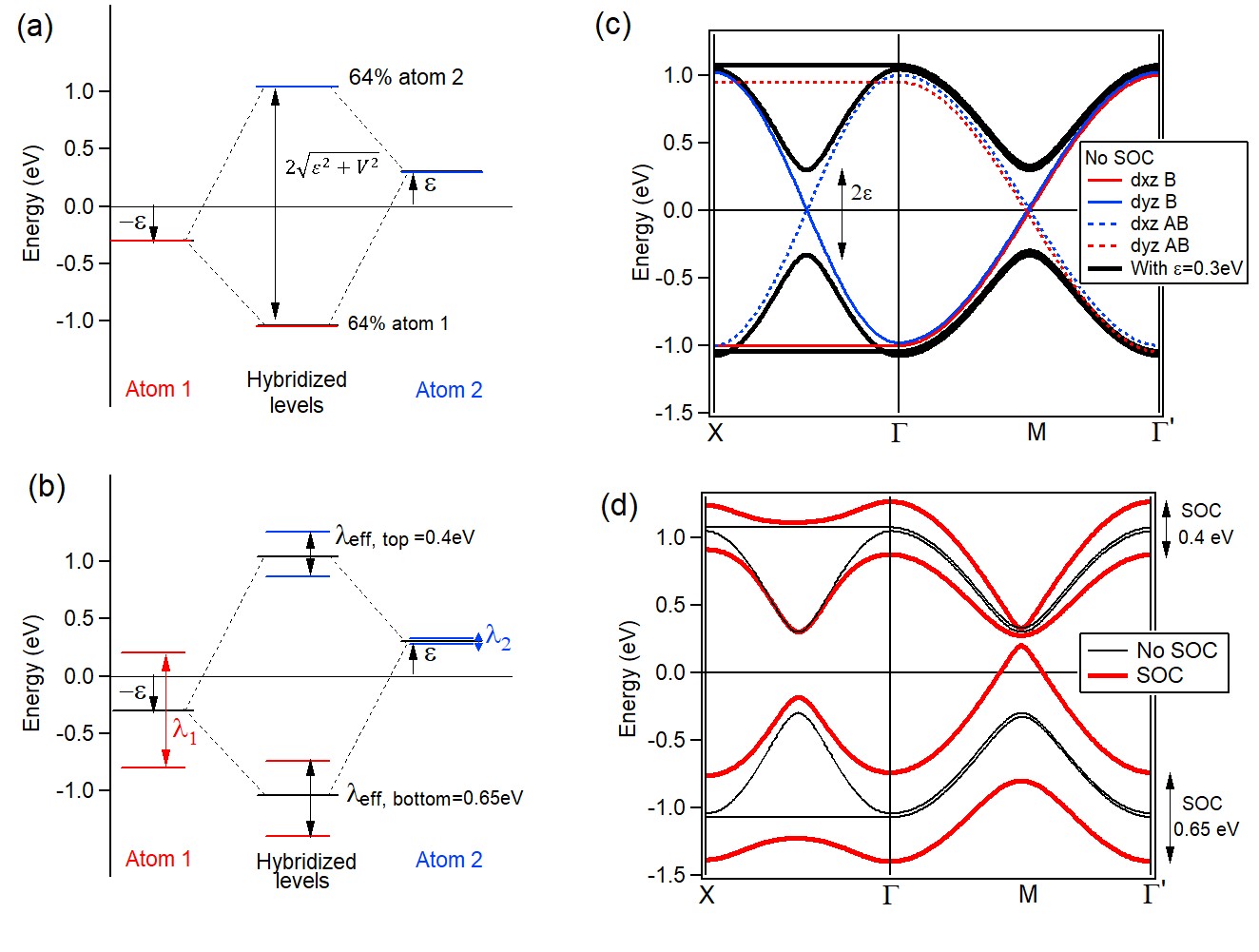}
\caption{(a) Simulation of the interaction between 2 energy levels distant from 2$\varepsilon$ and coupled by V. We used V=1eV, $\varepsilon$=0.3eV. (b) Same with SOC coupling between two degenerate energy levels $d_{xz}$ and $d_{yz}$ for each atom. The two atoms are characterized by different SOC values $\lambda_1$=1eV and $\lambda_2$=0.05eV. The new levels are split by an effective SOC of respectively 0.4eV (top) and 0.65eV (bottom). (c) Simulation of the band structure without SOC for two bands $d_{xz}$ and $d_{yz}$ with a cosine dispersion along $k_x$ and $k_y$. We consider a square lattice with two inequivalent atoms, such as the 2 Ir of \214 (in this case, $\varepsilon$=0, see red and blue bands) or ordered Ir and another TM (in this case, $\varepsilon$ can be different from zero, see black bands). (d) Same with SOC splitting, for the same parameters as (b), except that the interaction is k-dependent.} 
\label{SupTB}
\end{figure}
\vspace{0.2cm}
Although this model gives an interesting way to understand the different splitting at the Fermi level with Ru and Rh dopings, it is very difficult to push it further to get quantitative estimates. The parameters chosen here for the band width and $\varepsilon$=0.3eV are rather realistic for iridates and indeed give relative atomic contribution at the Fermi level (64\%-36\%) close to the ones obtained in the DFT calculation in Fig. 4. Using $\lambda$=0.5eV for Ir and $\lambda$=0.1eV for Ru and Rh, we would get two extreme values of effective $\lambda$, 0.35eV and 0.25eV. The modulation is significant, but does not easily explain the shift of 0.5eV of \L~between Rh and Ru. This emphasizes how the SOC value in iridates strongly deviates from this one electron picture. It is sensitive to correlations in many ways. First, correlations enhance the apparent SOC, by more than a factor 2 \cite{ZhouPRX18}. Second, the relative positions of the bands will change if the Mott gap closes, obviously changing the expected splitting. Third, the splitting could be renormalized if correlations induce renormalization as they do in Sr$_2$RuO$_4$ and Sr$_2$RhO$_4$. As these effects can all be different for Ir and Ru, it is really a different task to include them within this picture and we leave it for future studies.  

\vspace{0.3cm}

This model is very similar to the one used by Ref. \onlinecite{ZwartsenbergNatPhys20}, which recognized the importance of $\varepsilon$ in modifying the apparent SOC. In addition, they made calculations for disordered clusters, which allows to change the substitution value and avoid gaps due to artificially ordered structures. Their evaluation of $\lambda$, either in their calculation (see their Fig. 3) or their measurements (see their Fig. 4), is more sensitive to the Ir dominated bands, which is in our opinion the reason why they find a larger value of $\lambda$ for Ru than Rh at the same doping. While we agree on this value for the high binding energies, we argue that it is opposite at the Fermi level, $\lambda$ is smaller for Ru than Rh at the same doping, in agreement with the observed Fermi Surface structure. As it is the SOC value near E$_F$ that is relevant for the MIT, epecially through a modification of the filling of the bands, the SOC value cannot explain the different threshold for the MIT between Rh and Ru.

\vspace{0.5cm}
\noindent
\textbf{Hybridation}

In Fig. \ref{SupHyb}a, we show the DOS spectra for the three compounds, using their experimental structure at 300K. The structures are slightly different, there is no oxygen rotation in Sr$_2$RuO$_4$ and the Ir-O distance in the plane increases from 3.87\AA~(Ru) to 3.92\AA~(Rh) to 3.96\AA~(Ir). To evaluate the role of different structural parameters, we performed similar calculations using fictitious structures without oxygen rotation and different Ir-O distances. The changes are not negligible (typically 0.2eV shifts can be observed changing the Ir-O distance by 0.1\AA), but they do not change the overall trend. Moreover, the structure itself results from the strength of the hybridation between the transition metal and the oxygen, so that it appears more meaningful to compare the compound in their experimental structure. 

\begin{figure} [thb]
\centering
\includegraphics[width=0.9\linewidth]{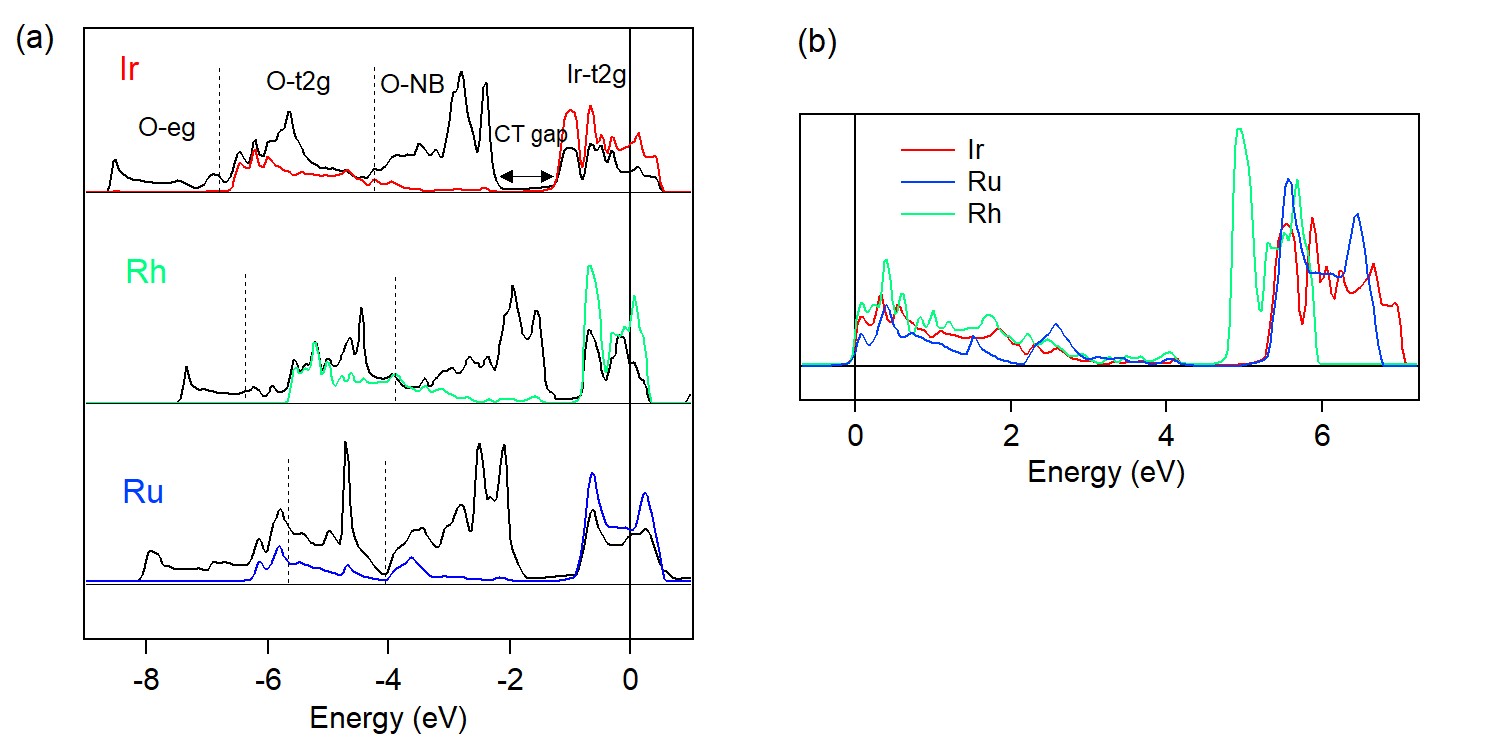}
\caption{(a) Total DOS (black line) calculated for the pure compounds : Sr$_2$IrO$_4$, Sr$_2$RuO$_4$, Sr2RhO4, with the structures observed experimentally. The conduction band is derived principally from the transition metal $t_{2g}$ states. The valence band is derived from oxygen bands, either non-bonding (NB), bonding with $t_{2g}$ (O-t2g) or with $e_g$ (O-eg), as indicated on the graph. The color line gives the DOS projected on the $d_{xz}/d_{yz}$  transition metal orbitals, which emphasizes the parts that are hybridized together. (b) The partial $d_{xz}/d_{yz}$ DOS of pannel (a) are aligned on the lower edge of the oxygen peak to evidence differences in hybridization. The intensity is normalized to the area of the $t_{2g}$ peak. The fact that the part in the oxygen valence band decreases in Ru compared to Ir and Rh indicates a larger difference $\varepsilon$ between O and Ru before hybridization.} 
\label{SupHyb}
\end{figure}

In all cases, the structure of the DOS is very similar, with the transition metal $t_{2g}$ contribution dominating at the Fermi level and the oxygen valence band between -2 and -8 eV displaying 3 peaks corresponding to non-bonding oxygens, bonding states hybridized with $t_{2g}$ and bonding states hybridized with $e_g$, as indicated. From this, it can be deduced that the splitting between bonding and antibonding states is approximately 5eV for $t_{2g}$ states and larger for $e_g$ states, as expected ($\simeq$9eV, AB $e_g$ states are located above 2eV). We show with color the partial contribution of $d_{xz}/d_{yz}$ orbitals, normalized to the intensity of the $t_{2g}$ peaks, which is an efficient way to locate the $t_{2g}$ contribution in bonding and antibonding peaks.

It is clear that the valence bands for Rh are much closer from the conduction bands than in the case of Ir, yielding a much smaller Charge Transfer (CT) gap. This reflects a smaller hybridization between Rh and Ir that we assign primarily to the different extension of $4d$ and $5d$ orbitals. However, the CT gap increases again for Ru, almost to the Ir value. We assign this to a difference in $\varepsilon$, due to the different electronegativity of Ru compared to Rh. As Ru has one less electron, therefore one less charge in the nucleus to attract electrons, Ru$^{4+}$ is not as well stabilized as O$^{2-}$ than Rh$^{4+}$ by Madelung energy, which means that its level before hybridation shall be higher compared to oxygen. This mechanism yields the well known tendency of smaller CT gap for late than early transition metals. In Fig. \ref{SupHyb}b, we align all spectra to the well defined lower edge of the $d_{xz}/d_{yz}$ DOS and indeed find that the distance to the t$_{2g}$ lower edge is smaller for Rh (4.7eV) than for Ir and Ru, which are almost identical around 5.4eV. It is also interesting to note that the contribution of the $d_{xz}/d_{yz}$ bands to the oxygen valence band is smaller for Ru than Rh and Ir (the ratio between the weight in bonding and antibonding bands is respectively : 0.5 (Ru), 0.8 (Rh) and 0.63 (Ir)). This confirms the idea that the Ru level is further away from O than Ir and Rh, applying the model described in the previous section (Fig. \ref{SupTB}a). 

We take these relative values as a basis for the sketch presented in Fig. 4. It is clear it is only a trend and the reality is much more complex : the structure would be different in the mixed compound, the valence might vary, the spin-orbit, which is not included, here would redistribute the states and, most importantly, correlation effects are neglected. Nevertheless, we believe that it identifies an important parameter for the different behavior of Ru and Rh that was mostly missing so far. More importantly, it allows to anticipate the importance of structural and doping effects in iridates

\end{widetext}

\end{document}